\documentclass[letter]{aa}

\usepackage{graphicx,dblfloatfix,fixltx2e}
\usepackage{txfonts}
\usepackage{hyperref}

\begin{document}

   \title{Anisotropy of the solar network magnetic field \\ around the average supergranule}

     \author{J. Langfellner \inst{1}
            \and
            L. Gizon \inst{2,1}
            \and
            A.~C. Birch \inst{2}}

    \institute{Georg-August-Universit\"at, Institut f\"ur Astrophysik,
               Friedrich-Hund-Platz 1, 37077 G\"ottingen, Germany
          \and
             Max-Planck-Institut f\"ur Sonnensystemforschung,
             Justus-von-Liebig-Weg 3, 37077 G\"ottingen, Germany
             }

   \date{Received <date> / Accepted <date>}

  \abstract
   {Supergranules in the quiet Sun are outlined by a web-like structure of enhanced magnetic field strength, the so-called magnetic network.
We aim to map the magnetic network field around the average supergranule near disk center.
We use observations of the line-of-sight component of the magnetic field from the Helioseismic and Magnetic Imager (HMI) onboard the Solar Dynamics Observatory (SDO).
The average supergranule is constructed by coaligning and averaging over 3\,000 individual supergranules. We determine the positions of the supergranules with an image segmentation algorithm that we apply on maps of the horizontal flow divergence measured using time-distance helioseismology.
In the center of the average supergranule the magnetic (intranetwork) field is weaker by about 2.2~Gauss than the background value (3.5~Gauss), whereas it is enhanced in the surrounding ring of horizontal inflows (by about 0.6~Gauss on average). We find that this network field is significantly stronger west (prograde) of the average supergranule than in the east (by about 0.3~Gauss). With time-distance helioseismology, we find a similar anisotropy.
The observed anisotropy of the magnetic field adds to the mysterious dynamical properties of solar supergranulation.}

   \keywords{Sun: magnetic fields -- Sun: helioseismology -- Convection}

   \maketitle


\section{Introduction}
Solar supergranules are surrounded by the network magnetic field that can be observed, for instance, in \ion{Ca}{ii} K emission lines in the solar chromosphere \citep[e.g.,][]{simon_1964}. The network field is built up through the advection of magnetic field by supergranular flows \citep[e.g.,][]{rieutord_2010}.
Beyond this, however, not much is known about the dynamical interaction of supergranulation and the network field. The dynamics of supergranulation itself is not understood \citep[e.g.,][]{gizon_2003,rieutord_2010}.

In this letter, we present photospheric maps of the magnetic field of the average supergranule using data from the Helioseismic and Magnetic Imager (HMI) \citep{schou_2012} onboard the Solar Dynamics Observatory (SDO) at full resolution (about 1~arcsec). The average supergranule is constructed as an ensemble average of individual supergranules that are identified in maps of the horizontal flow divergence from time-distance helioseismology \citep{duvall_2000}.


\section{Observations and data processing} \label{sect_observations}
We tracked $12\degr\times 12\degr$ patches of HMI line-of-sight magnetograms near disk center from 1 May through 28 August 2010, using a tracking rotation rate of $14.33\degr$ per day. The patches were remapped using Postel's projection and a spatial sampling of 0.348~Mm. The temporal cadence is 45~s. The 24-h datacubes are centered around the central meridian. Each 24-h datacube is divided in three 8-h chunks. Note that the line-of-sight magnetic field $B_\text{LOS}$ near disk center consists almost purely of the vertical magnetic field component.

In the same way and for the same patches, we tracked and remapped HMI line-of-sight Dopplergrams. 
We apply f-mode time-distance helioseismology to the 8-h datacubes to infer the horizontal divergence of the flow field \citep{langfellner_2014,langfellner_2015}. This is achieved by computing the temporal cross-correlation between each point and a surrounding annulus (10~Mm radius) and measuring the outward minus inward travel times. 
Additionally, we computed the average of inward and outward travel times, i.e.~the mean travel times. The mean travel times are known to be sensitive to the magnetic field \citep{duvall_2006}.

   \begin{figure*}[h]
\centering
\includegraphics[width=0.5\hsize]{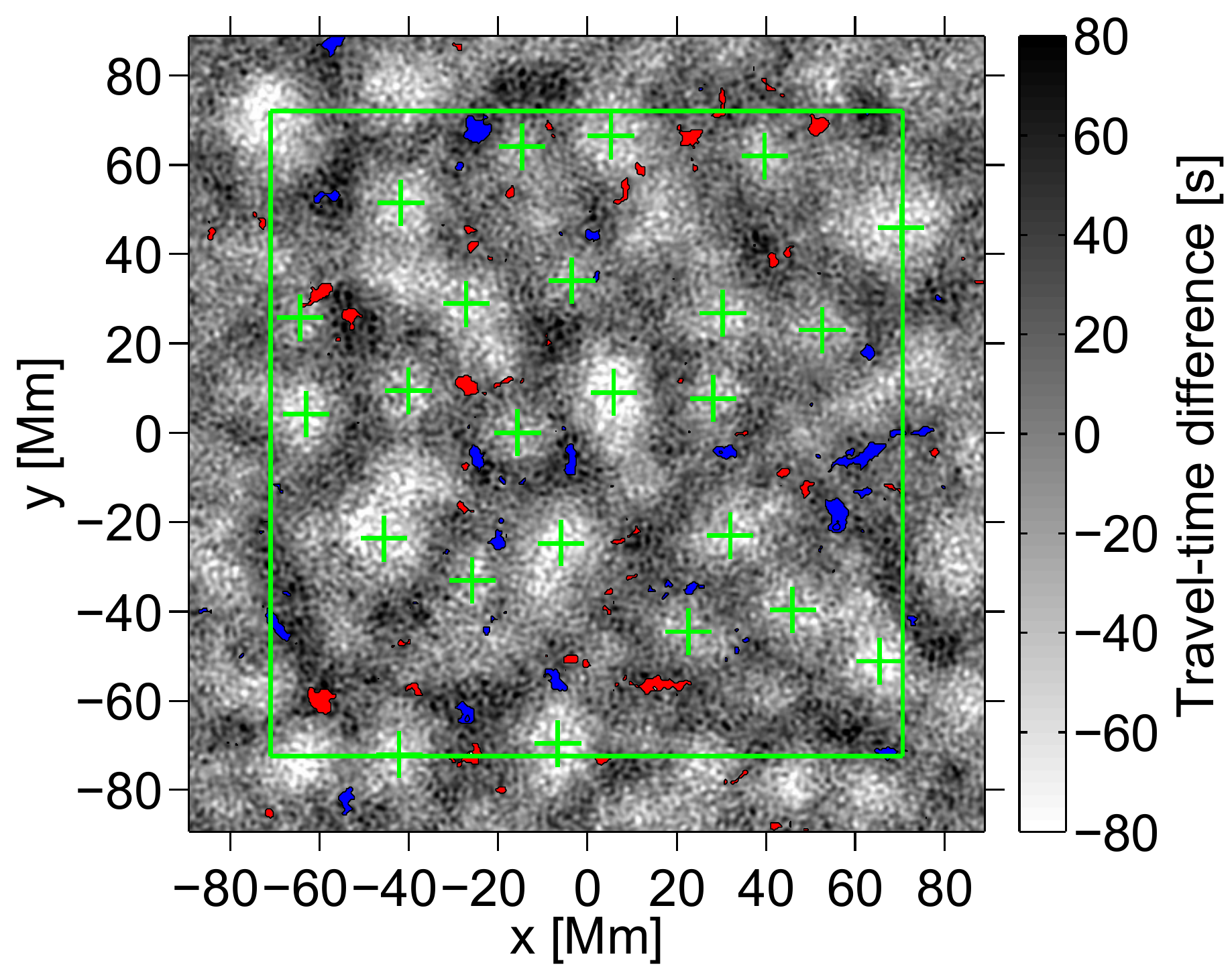}
\includegraphics[width=0.49\hsize]{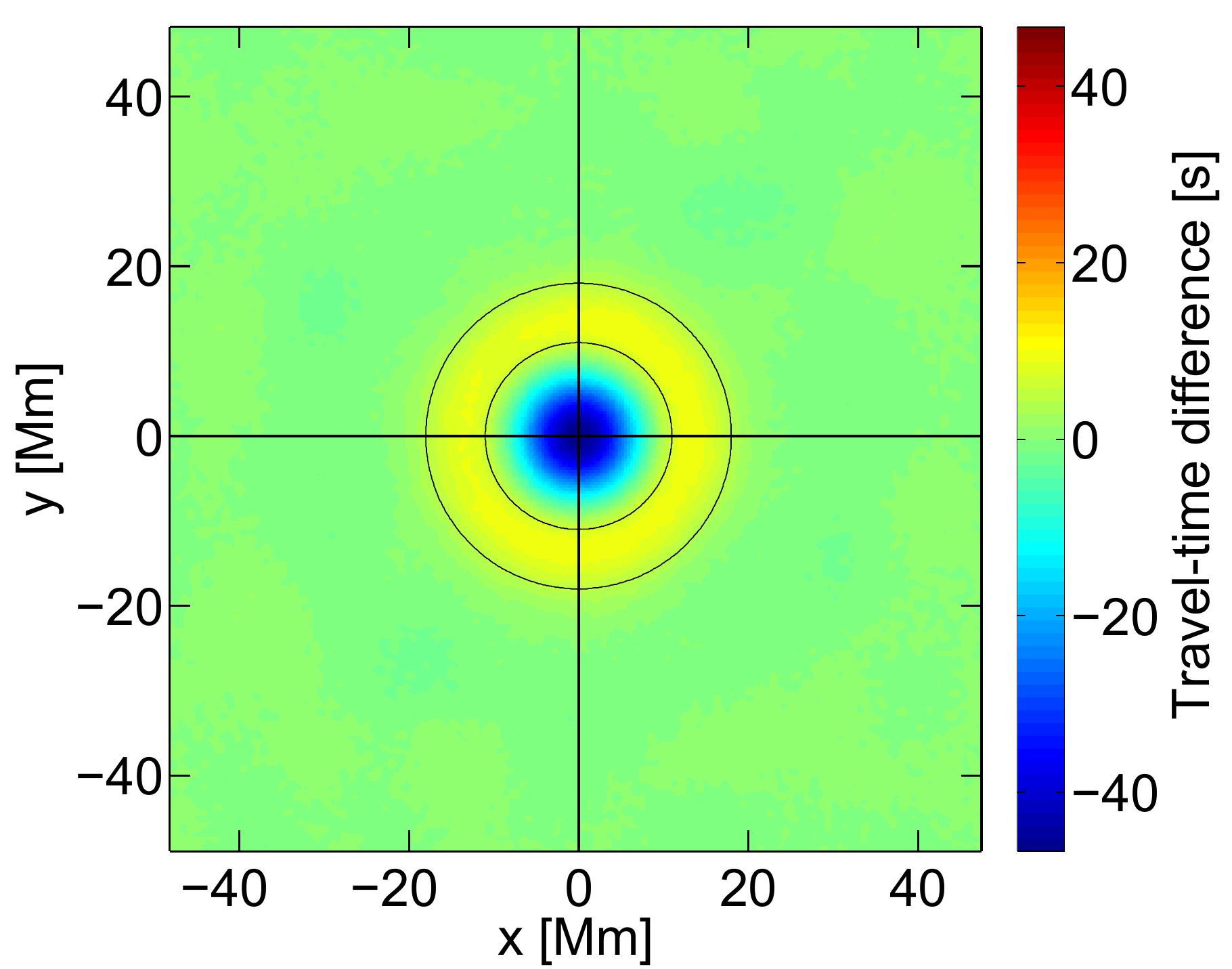}
\caption{\textbf{Left}: Positions (green crosses) of centers of supergranules identified near disk center on 2 May 2010. Supergranules that are outside of the green frame have been discarded. The greyscale image shows the f-mode travel-time differences (white is for outflows, black for inflows) using 8~h of SDO/HMI observations. The colorbar is truncated in the range between $-80$ and 80~s.  The filled contours mark areas where the HMI line-of-sight magnetic field averaged over 8~h exceeds $20$~Gauss in absolute value (red is positive field strength, blue is negative). \textbf{Right:} Travel-time map for the average supergranule, where negative (positive) values correspond to horizontal outflows (inflows). The black circles have radii of 11 and 18~Mm.}
\label{fig1}
    \end{figure*}

From the divergence maps, we identified the supergranule boundaries using the image segmentation algorithm from \citet{hirzberger_2008}. The left panel of Fig.~\ref{fig1} shows an example 8-h divergence map with magnetic field contours overlaid. The centers of supergranules (at maximum divergence) are marked with green crosses. Following \citet{duvall_2010} and \citet{langfellner_2015}, we construct an average supergranule by shifting and averaging over all the individual supergranules (about 3\,000). In the right panel of Fig.~\ref{fig1}, we show the horizontal divergence of the average supergranule. The average outflow is surrounded by a ring of inflows with a radius of 15~Mm.


\section{Results}

\subsection{Magnetic field of the average supergranule near disk center}

   \begin{figure*}[h]
  \centering
\includegraphics[width=0.48\hsize]{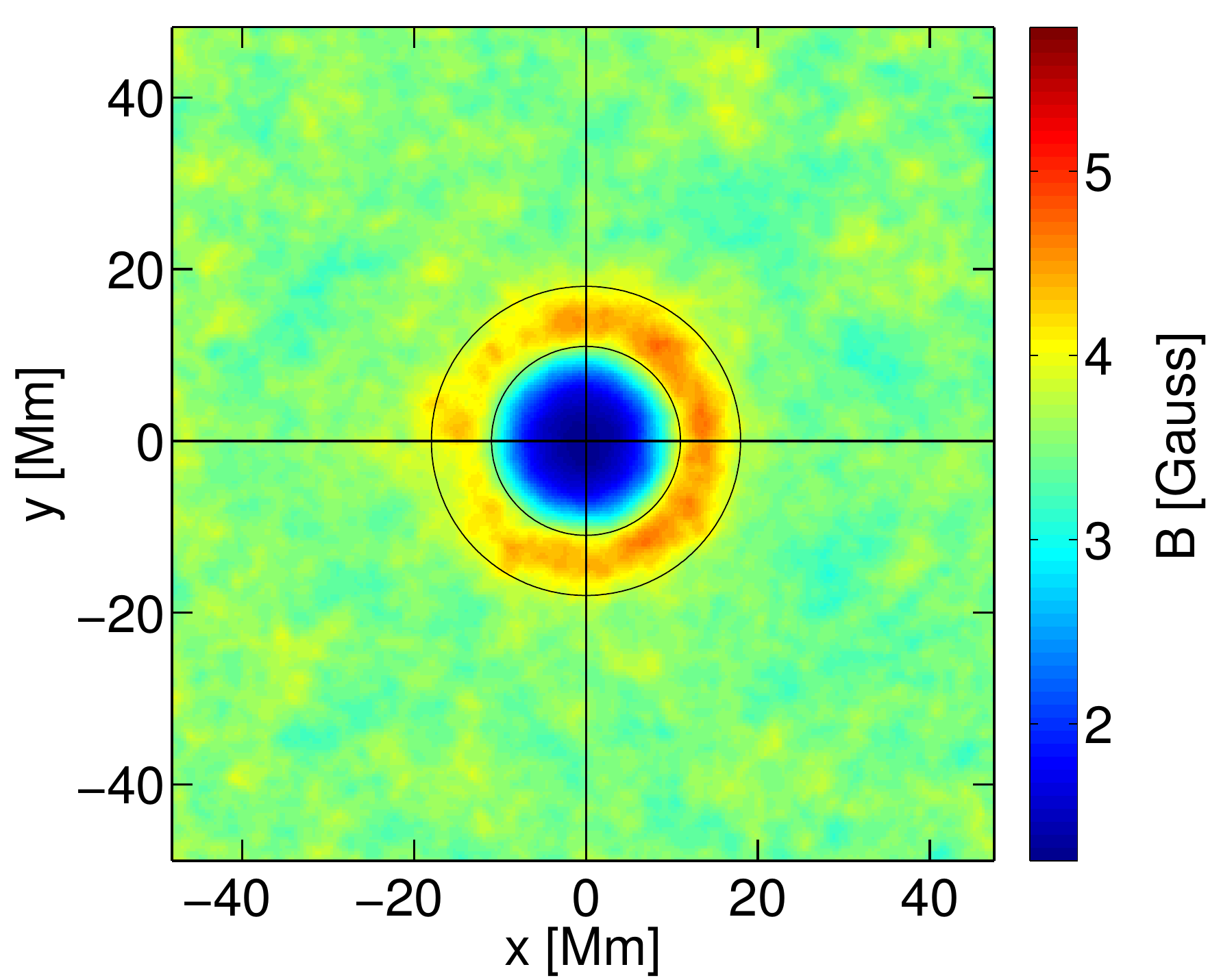}
\includegraphics[width=0.495\hsize]{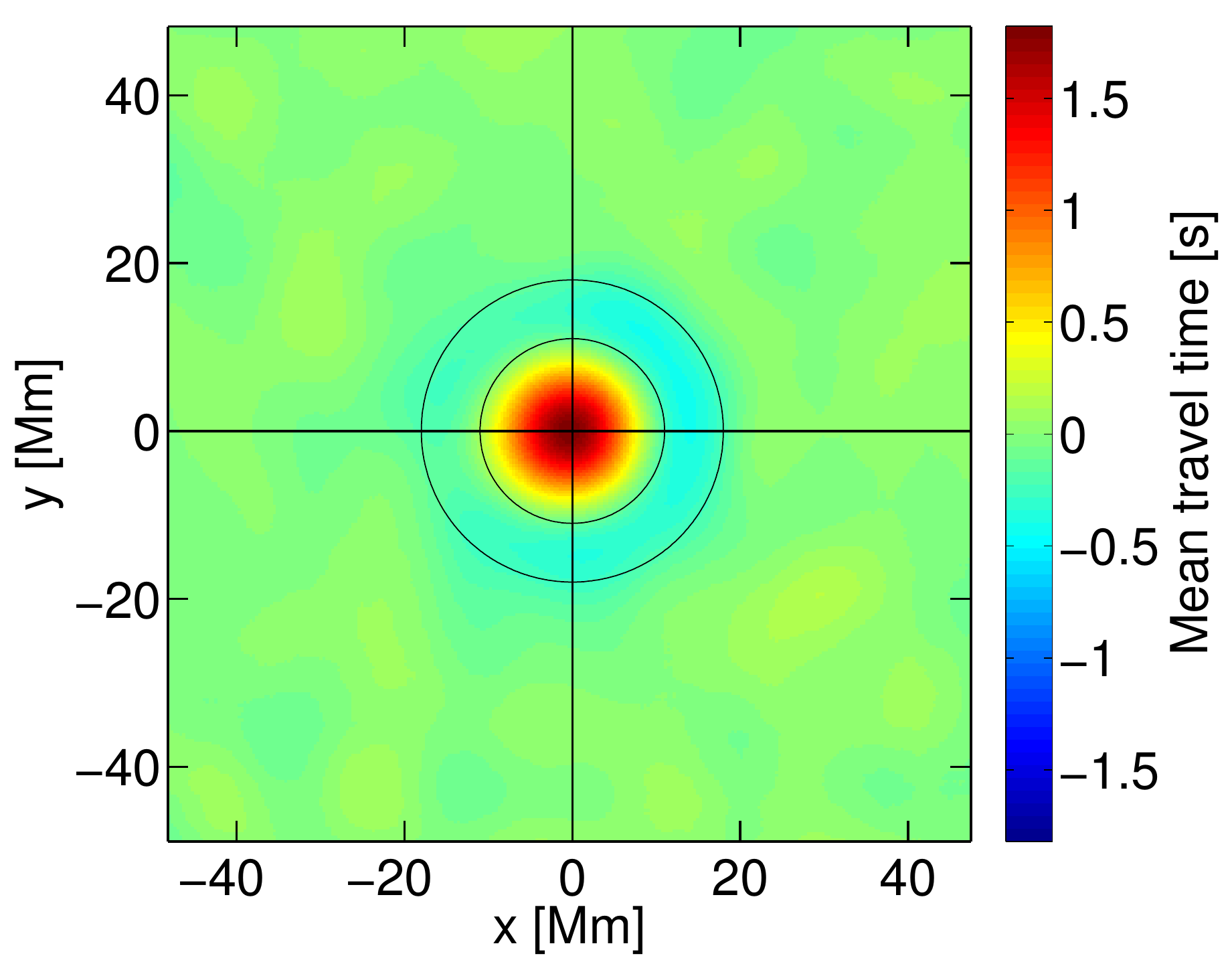}
\caption{Line-of-sight magnetic field and mean travel times of the average supergranule near disk center, as measured from HMI. The black circles are centered around the origin and have radii of 11 and 18~Mm. \textbf{Left:} Absolute line-of-sight magnetic field $B$. The absolute value was taken after averaging $B_\text{LOS}$ over 8~h. The colorbar is symmetrized around the spatial average ($3.52$~Gauss). \textbf{Right:} Mean travel times for f modes (surface gravity waves). These travel times are the mean of outward and inward travel times measured between a central point and a surrounding annulus of radius 10~Mm and are sensitive to the magnetic field. The colorbar is symmetrized around zero (which is the mean value by construction).}
\label{fig2}
    \end{figure*}

By using the coordinates of the supergranules determined in the divergence maps, we 
construct a magnetic field map for the average supergranule (left panel of Fig.~\ref{fig2}).
The quantity we average over is the absolute value of $B_\text{LOS}$, where the absolute value has been taken after averaging $B_\text{LOS}$ over the length of a datacube (8~h). We use $B$ to denote this quantity.
Rapidly varying small-scale magnetic field is substantially suppressed in this analysis.
The spatially averaged $B$ has a value of $\overline{B}=3.52$~Gauss. In the center of the average supergranule, the magnetic field is weaker than $\overline{B}$ by about $2.2$~Gauss. In the inflow region, on the other hand, the (network) magnetic field is stronger than the average value by up to roughly 1~Gauss. 
Surprisingly, the network field of the average supergranule is stronger in the west (in the prograde direction) than in the east. This is a statistically highly significant result (as we will show later).

A similar anisotropy is observed in the mean travel times (right panel), with a larger amplitude. In the center of the supergranule, the mean travel times are larger than in the inflow region where the magnetic field is stronger. In comparison to the divergence  map in Fig.~\ref{fig1}, the central peak of the mean travel time is slightly shifted to the east. Note that the peak amplitude of the mean travel times \citep[presumably caused by magnetic field, e.g.,][]{duvall_2006} is smaller by a factor of 25 compared to the peak in the travel-time differences \citep[mostly caused by radial outflows with a peak velocity of about 300~m~s$^{-1}$,][]{langfellner_2015}.

   \begin{figure*}
  \centering
\includegraphics[width=0.48\hsize]{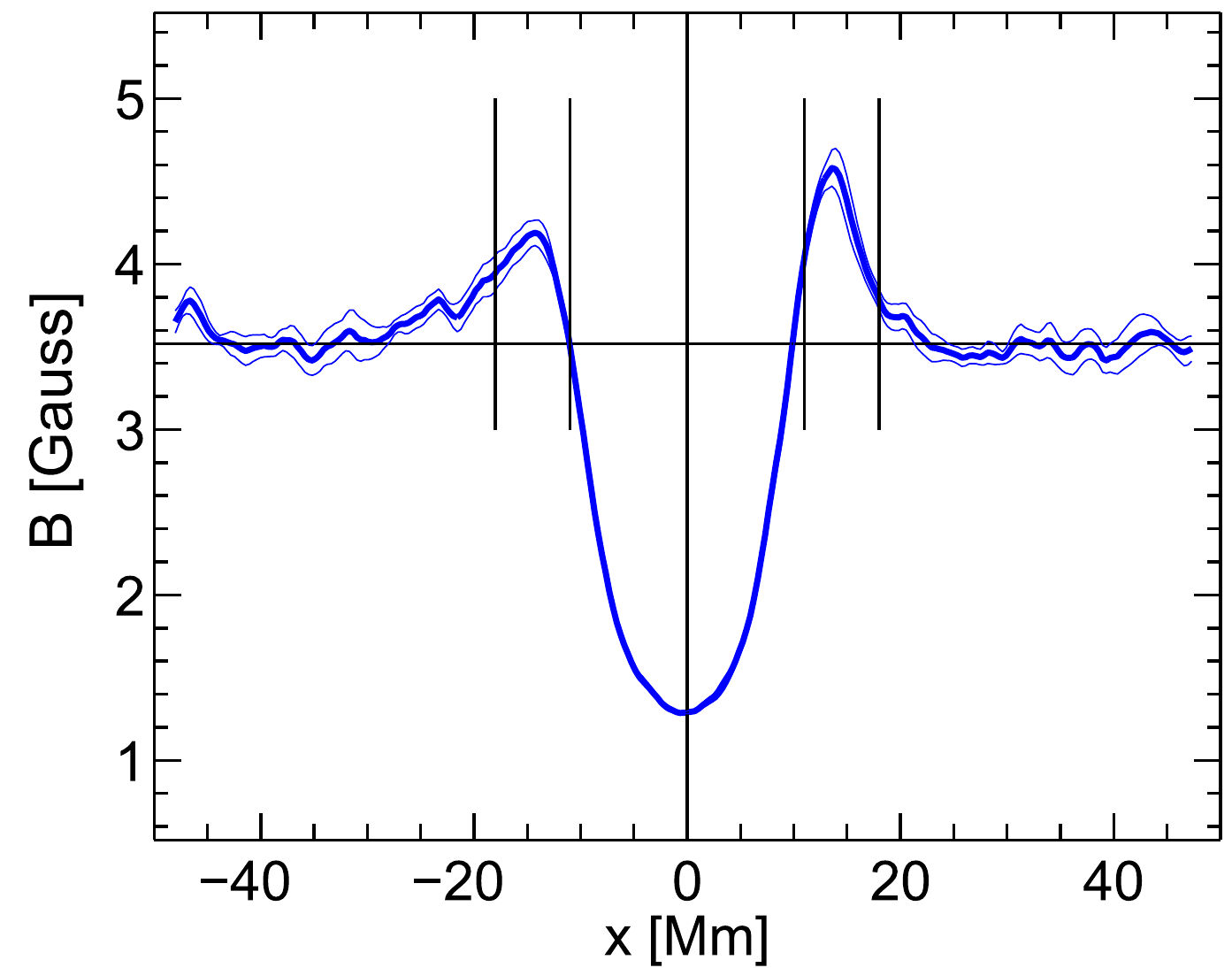}
\includegraphics[width=0.472\hsize]{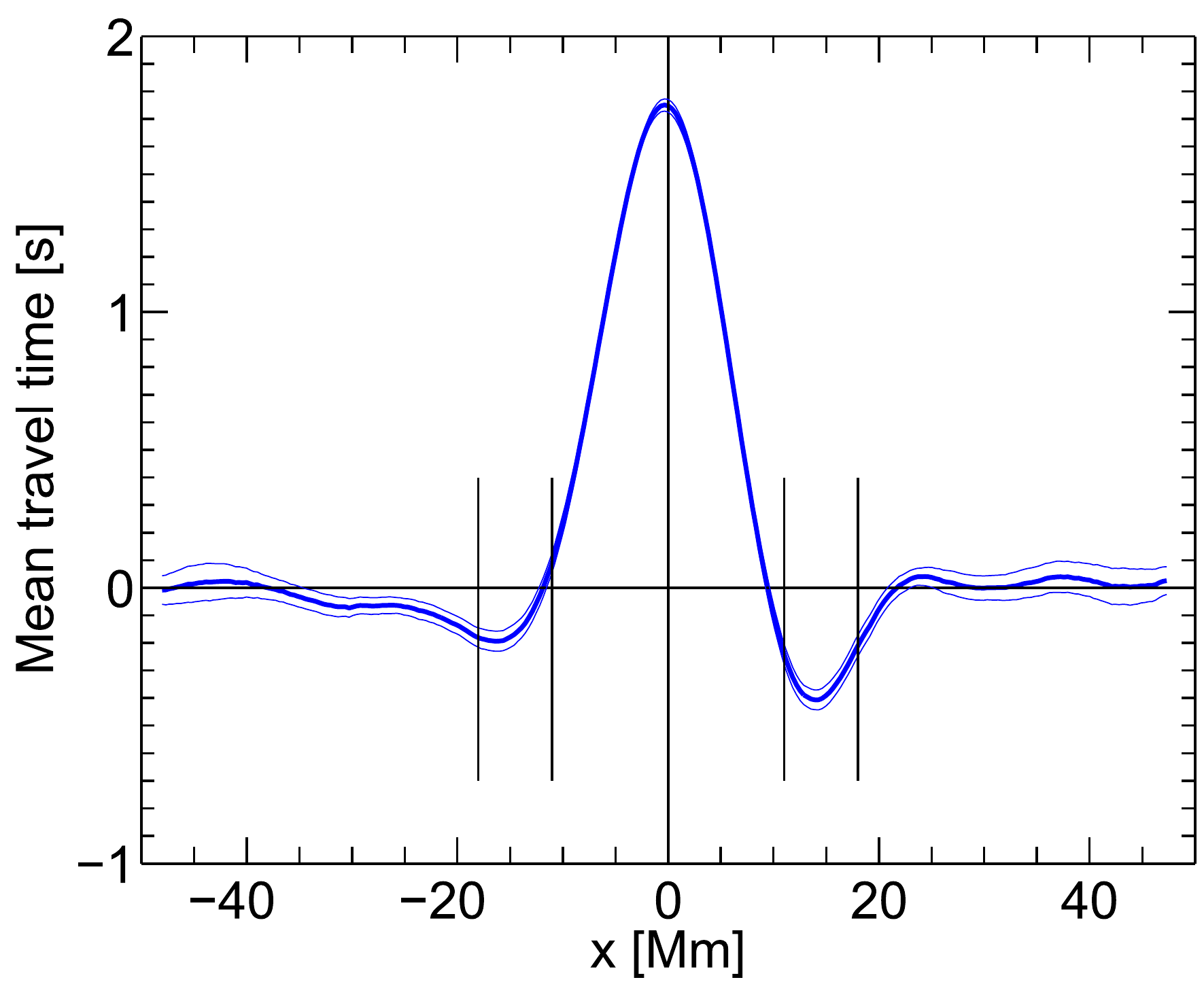}
\caption{Cuts along $x$ axis (averaged over a strip $|y| < 2.5~$Mm) of the panels in Fig.~\ref{fig2}. The blue thin lines give the $1$-$\sigma$ bounds, the vertical lines indicate the position of the ring (network) shown in the previous plots. \textbf{Left:} Line-of-sight magnetic field, $B$. The horizontal black line marks the spatial average ($3.52~$Gauss). \textbf{Right:} Mean travel times for f modes.}
\label{fig3}
    \end{figure*}

Figure~\ref{fig3} shows plots along the $x$-axis through the maps from Fig.~\ref{fig2}, after averaging over a band $|y| < 2.5~$Mm. The dip of the magnetic field in the center of the supergranule is rather flat compared to a Gaussian profile and has a full width at half maximum (FWHM) of about 16~Mm. On the west side of the surrounding ring, the field is about $0.3$ to $0.4$~Gauss stronger than on the east side (this difference corresponds to more than $3\sigma$). The maximum field is attained at $x = \pm13$~Mm.
For the mean travel times, the FWHM of the central peak is about 12~Mm and more Gaussian in shape. The zero-crossing positions are found at $x=9$ and $x=-12$~Mm. Furthermore, the minima are at different distances (13 and 15~Mm). The mean travel time on the west side has about twice the magnitude than on the east side.

Our findings can be compared to \citet{duvall_2010} who measured the absolute line-of-sight magnetic field of the average supergranule using data from the Michelson Doppler Imager (MDI) \citep{scherrer_1995} onboard the Solar and Heliospheric Observatory (SOHO), albeit for the azimuthally averaged $B$. Their profile of $B$ as a function of the distance from the center of the average supergranule agrees with our measurements but there are differences in the details of the curves. \citeauthor{duvall_2010} measured a small bump in the central dip (perhaps not significant) and a maximum located at a distance of about 18~Mm to the supergranule center, thus further away than in our measurements (13~Mm). This is probably due to a different selection of supergranulation sizes. Their magnetic field is about 2~Gauss in the dip and has a maximum of $5.5$~Gauss. Their average field is roughly 4~Gauss, close to our value of $3.52$~Gauss. Note that \citeauthor{duvall_2010} averaged only over 4~h per map though, which could explain their stronger average field.

\subsection{Measuring the anisotropy of the network field}
In order to analyze the anisotropy of the network field in more detail, we study $B$ in the network as a function of azimuthal angle $\psi$, where $\psi=0\degr$ points west and $\psi=90\degr$ points north. This is accomplished by averaging $B$ between the circles with radii $11$ and $18$~Mm shown in Fig.~\ref{fig2} over azimuthal segments of width $10\degr$. The result we call $B_\text{network}$. We apply the same procedure to the mean travel times.

   \begin{figure*}
  \centering
\includegraphics[width=0.495\hsize]{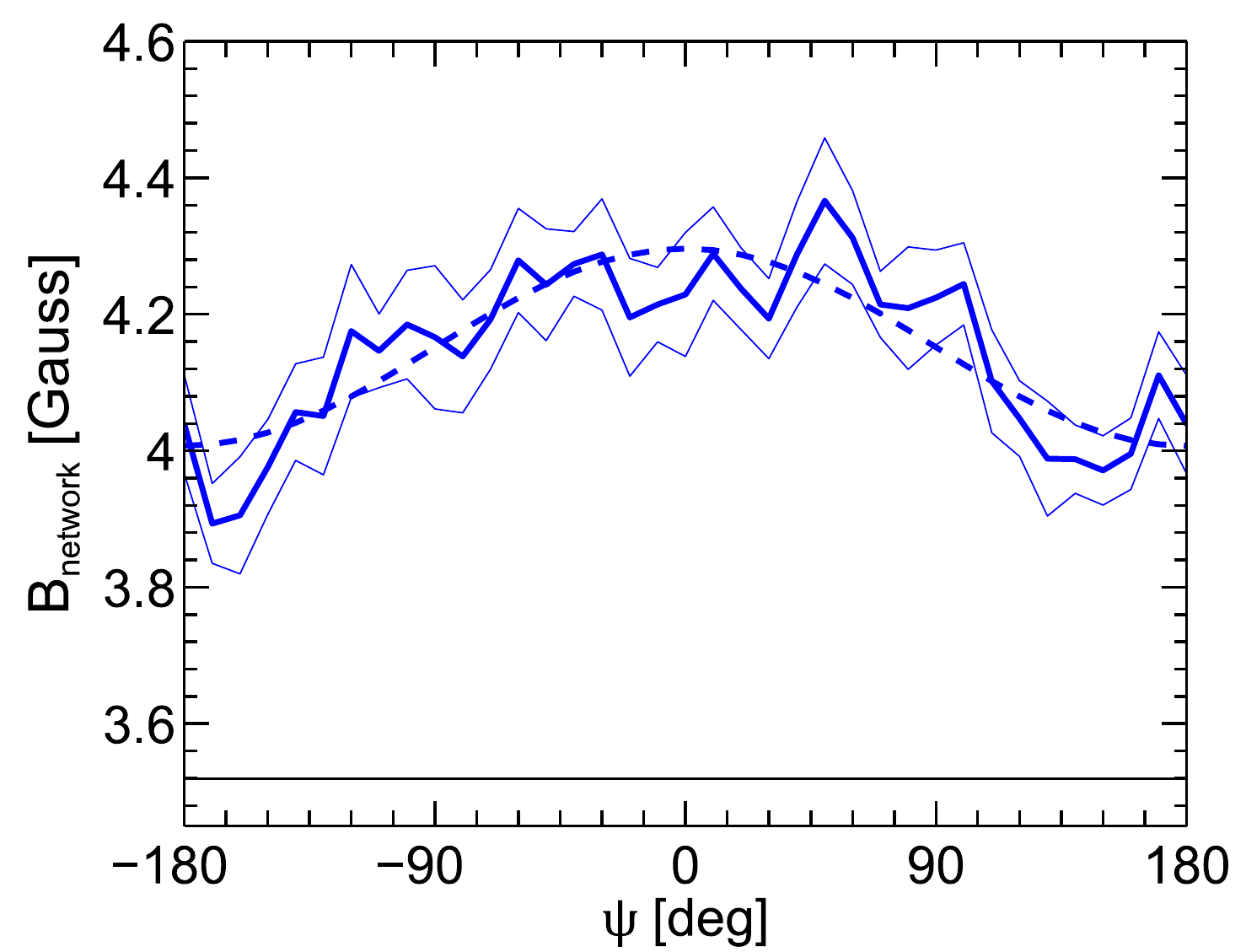}
\includegraphics[width=0.495\hsize]{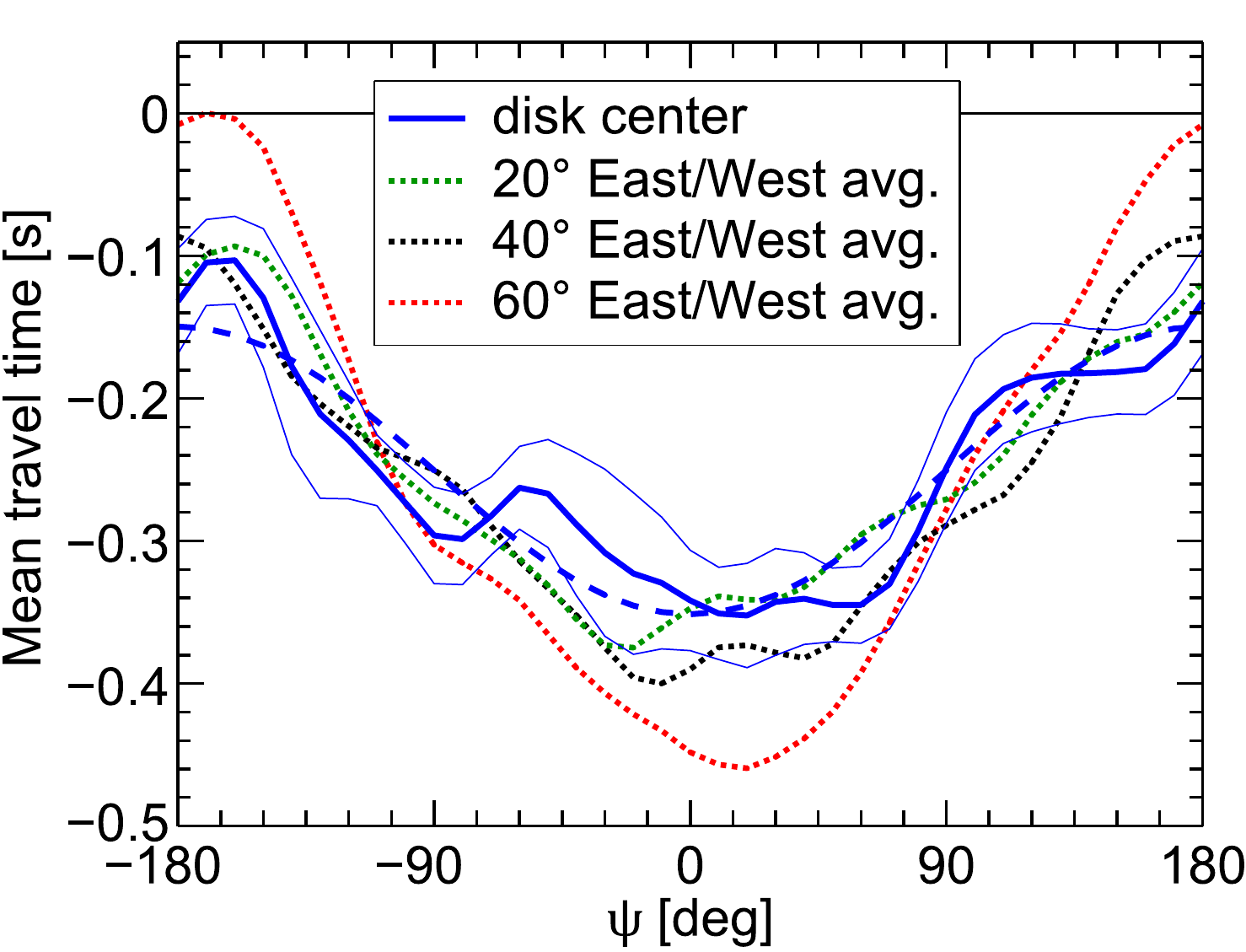}
\caption{\textbf{Left:} Magnetic field $B_\text{network}$ around the average supergranule near disk center as a function of azimuthal angle $\psi$. To obtain $B_\text{network}$, the magnetic field $B$ was averaged over segments (width $10\degr$ in $\psi$) of a concentric ring around the center of the average supergranule. The inner and outer limits of the ring are 11 and 18~Mm (circles in Fig.~\ref{fig2}). The thin blue lines give the $1$-$\sigma$ bounds. The dashed line is a least-squares fit to a cosine of the form specified by Eq.~(\ref{eq_network}). The maximum of the cosine is fixed at $\psi=0\degr$ (west direction). For comparison, the horizontal black line shows the background field $\overline{B} = 3.52~$Gauss. \textbf{Right:} As left, but for f-mode mean travel times. In addition to the result at disk center, the mean travel times at other positions along the equator are shown (using patches centered at longitudes $\pm20\degr$, $\pm40\degr$, and $\pm60\degr$ away from the central meridian, averaged over east and west). The dashed line represents the cosine function given by Eq.~(\ref{eq_network_tt}). By construction, the mean travel time in the background is zero.}
\label{fig4}
    \end{figure*}

The azimuthal dependence of $B_\text{network}$ is shown in Fig.~\ref{fig4}. As already seen in Fig.~\ref{fig2}, $B_\text{network}$ is significantly larger near $\psi = 0\degr$ (in the west direction) than near $\psi = 180\degr$ (east direction). We find that the azimuthal dependence of $B_\text{network}$ can be described by 
\begin{equation}
B_\text{network} \approx 3.52 + 0.63 (1 + 0.23 \cos \psi) ~\text{Gauss} .  \label{eq_network}
\end{equation}
 This means that $B_\text{network}$ varies from $4.0$~Gauss in the east to $4.3$~Gauss in the west.

For the mean travel times (right panel of Fig.~\ref{fig4}), the situation is similar, but the variations are of opposite sign: 
\begin{equation}
\delta \tau_{\rm network} \approx   -0.25  (1 + 0.40 \cos \psi) ~{\rm s} . \label{eq_network_tt}
\end{equation}

One may ask whether the anisotropy in the mean travel times is caused by instrumental errors. We believe that this is unlikely. First, the HMI instrument shows no astigmatism for any practical purposes (in particular, the power of solar oscillations is isotropic at disk center). Second, we find that the anisotropy in the travel times around the average supergranule is the same for patches along the equator but centered at longitudes $\pm20\degr$, $\pm40\degr$, and $\pm60\degr$ away from the central meridian, averaged over east and west.

Because it is known that f-mode travel times are reduced in magnetic regions \citep[e.g.,][]{gizon_2006,duvall_2006}, the two results for $B_\text{network}$ and $\delta \tau_{\rm network}$ appear to be consistent.
We note that the relative variations of the travel times ($0.40$) are larger than the relative variations of the magnetic field ($0.23$).
Furthermore, by considering all the points in the average-supergranule maps, we find a one-to-one relationship between $\delta\tau$ and $B$. By choosing the center of the average supergranule as the reference point with $B_\text{min} = 1.3~$Gauss and $\delta \tau_\text{max} = 1.8~$s,
this relationship can be described by 
\begin{equation}
\delta\tau \sim \delta\tau_\text{max} - (1.2~\text{s Gauss}^{-1/2}) (B - B_\text{min})^{1/2}  .
\end{equation}
This suggests that the magnetic field anisotropy leaves its signature in the travel-time perturbations. However, it is not excluded that an anisotropy in the flows could also affect the travel times.


\section{Conclusion}
Using data from SDO/HMI, we have measured the line-of-sight component of the magnetic field of the average supergranule near disk center (ensemble average over about 3\,000 supergranules). We detected an unexpected anisotropy of the network field that surrounds the average supergranule. The magnetic field is stronger west of the average supergranule. Similarly, the mean travel time is decreased west of the average supergranule.

The measured anisotropy of the magnetic field is but one result in a series of puzzling observations associated with supergranulation. Other observations include, for example, the finding that the supergranulation pattern rotates faster around the Sun than magnetic features \citep[e.g.,][]{snodgrass_1990, meunier_2007} and the wavelike properties of supergranulation  \citep{gizon_2003,schou_2003}. How these phenomena are related to  the  discovery in this paper is unclear.

\begin{acknowledgements}
JL, LG, and ACB designed research. JL performed research, analyzed data, and wrote the paper. We thank T. Duvall and J. Schou for useful discussions.
JL and LG acknowledge research funding by Deutsche Forschungsgemeinschaft (DFG) under grant SFB 963/1 ``Astrophysical flow instabilities and turbulence'' (Project A1).
The HMI data used are courtesy of NASA/SDO and the HMI science team.
The data were processed at the German Data Center for SDO (GDC-SDO), funded by the German Aerospace Center (DLR). Support is acknowledged from the SPACEINN and SOLARNET projects of the European Union.
We are grateful to R. Burston and H. Schunker for providing help with the data processing, especially the tracking and mapping. We used the workflow management system Pegasus funded by The National Science Foundation under OCI SI2-SSI program grant \#1148515 and the OCI SDCI program grant \#0722019.
\end{acknowledgements}

\bibliographystyle{aa}
\bibliography{literature}

\Online

\end{document}